\documentclass[sn-mathphys-num]{sn-jnl}

\usepackage{graphicx}%
\usepackage{multirow}%
\usepackage{amsmath,amssymb,amsfonts}%
\usepackage{amsthm}%
\usepackage{mathrsfs}%
\usepackage[title]{appendix}%
\usepackage{xcolor}%
\usepackage{textcomp}%
\usepackage{manyfoot}%
\usepackage{booktabs}%

\usepackage{listings}%

\usepackage{amsmath,amsfonts,amsbsy,amssymb,mathtools,amsthm,tabulary,graphicx,times,fancyhdr, mathrsfs, pgf, epsfig, epstopdf, lscape,booktabs, enumitem,lipsum,tkz-graph,scalerel,stackengine,setspace,kbordermatrix,adjustbox,float,caption,subcaption}

\theoremstyle{thmstyleone}%
\theoremstyle{thmstyletwo}%

\theoremstyle{thmstylethree}%

\begin{document}

\title{Inferring Grain Size Distributions from Magnetic Hysteresis in M-type Hexaferrites}


\author*[1]{\fnm{Masoud} \sur{Ataei}}\email{masoud.ataei@utoronto.ca}
\author[2]{\fnm{Mohammad Jafar} \sur{Molaei}}
\author[3]{\fnm{Abolghasem} \sur{Ataie}}

\affil[1]{\normalsize\orgdiv{Department of Mathematical and Computational Sciences}, \orgname{University of Toronto}, \state{Ontario}, \country{Canada}}

\affil[2]{\normalsize\orgdiv{Faculty of Chemical and Materials Engineering}, \orgname{Shahrood University of Technology}, \state{Shahrood}, \country{Iran}}

\affil[3]{\normalsize\orgdiv{School of Metallurgy and Materials Engineering, College of Engineering}, \orgname{University of Tehran}, \state{Tehran}, \country{Iran}}



\abstract{
We develop a stochastic-dynamic framework to infer latent grain size distribution from magnetic hysteresis data in M-type hexaferrite materials, offering an alternative to imaging-based characterization. A stochastic nucleation-growth process yields a Modified Lognormal Power-law grain size distribution. This is combined with Brown's relation to obtain a coercivity probability distribution, which is embedded within a dynamic magnetization model. A key feature is the joint estimation of microstructural parameters, including the critical grain radius, through inverse optimization of full hysteresis loops. Experimental validation on hydrothermally synthesized strontium hexaferrite subjected to nitrogen treatment and recalcination reveals interpretable trajectories of nucleation, growth, and structural memory encoded in the magnetic response.
 }



\maketitle

\section{Introduction}

Magnetic hysteresis in polycrystalline hexaferrites is strongly modulated by the statistical distribution of grain sizes. Processes of nucleation, growth, and impingement that occur during synthesis and post-synthesis treatments leave lasting signatures on macroscopic magnetic behavior, particularly coercivity and magnetization reversal dynamics~\cite{cullity2011introduction, abbaschian2010physical}. In M-type strontium hexaferrite, coercivity is known to be especially sensitive to the underlying distribution of grain sizes, which in turn reflects the sample's thermal and chemical history~\cite{pullar2012hexagonal, Molaei2025}.

While electron microscopy offers direct access to grain-scale morphology, it suffers from inherent limitations in statistical representativeness and segmentation bias. Magnetic hysteresis measurements, in contrast, integrate the collective response of the entire grain ensemble and thereby encode latent information about microstructural variability. This motivates the development of inference methods that extract grain size statistics from hysteresis data, providing a non-invasive and statistically robust complement to imaging-based techniques.

In this work, we propose a modeling framework that supports such inference. Grain growth is modeled as a stochastic-dynamic process wherein deterministic expansion is modulated by stochastic arrest. The resulting grain size distribution is captured by the Modified Lognormal Power-law (MLP) model~\cite{basu2004power, basu2015mlp}, which accounts for both bounded growth and rare overgrowth events. Structural variability is then propagated into coercivity space via Brown's relation~\cite{cullity2011introduction,brown1959micromagnetics}, and embedded within a time-dependent magnetization model.

An important feature of the proposed approach is that the critical grain radius is treated as an unknown parameter, optimized jointly with the MLP distribution parameters and dynamic relaxation times. This enables the model to adapt across different processing regimes and reveal physically interpretable transitions in grain-scale behavior.

We validate the framework using experimental hysteresis data from hydrothermally synthesized strontium hexaferrite subjected to nitrogen heat treatment~\cite{Molaei2025,ataie1996heat}. Such treatments are known to induce transient phase decomposition and grain refinement, followed by recovery of the hexagonal structure via recalcination, a process governed by a robust \emph{structural memory effect}~\cite{Molaei2025}. The model accurately captures the essential features of hysteresis evolution across treatments, while yielding microstructural descriptors that align with independent imaging observations~\cite{ataie1996heat}.

The remainder of this paper is structured as follows. Section~\ref{Sec:Literature} provides an overview of relevant literature and prior modeling approaches. Section~\ref{sec:nucleation_growth} introduces the stochastic nucleation-growth model used to describe grain size distributions and their kinetic underpinnings. Section~\ref{sec:magnetic_modeling} presents the theoretical formulation of the coercivity and magnetization models, including dynamic extensions that account for time-dependent effects. Section~\ref{sec:experiments} describes the parameter estimation methodology and experimental conditions. This section further presents results across multiple processing stages, interprets the inferred microstructural parameters, and evaluates model predictions against empirical hysteresis data. Finally, Section~\ref{sec:conclusion} summarizes the main findings and outlines directions for further model refinement and extension.

\section{Related Work}
\label{Sec:Literature}

To situate the current development within the broader framework of stochastic modeling of grain size and magnetic properties, it is helpful to consider recent advances in the area. Probabilistic modeling of grain size evolution has a rich history across materials science. In crystallization and grain growth processes, stochastic nucleation and growth (RNG) models have been extensively developed to explain the emergence of characteristic grain size distributions. Analytical frameworks based on RNG kinetics capture the interplay of time-dependent nucleation, impingement-limited growth, and kinetic arrest, yielding grain size distributions including lognormal, truncated lognormal, and power-law forms~\citep{teran2010time,bergmann2002growth,kumomi1999alternative,bergmann1998formation,bolyachkin2018power,bance2014grain,lee2013particle}. In particular, Teran \emph{et al.}~\citep{teran2010time} demonstrated that a broad class of nucleation-growth processes leads naturally to truncated lognormal grain size distributions, even without assuming such a form \emph{a priori}, by accounting for the dynamic reduction of both effective nucleation and growth rates during impingement. Experimental studies similarly confirm the robustness of lognormal grain size distributions across diverse crystallization contexts~\citep{bergmann1998formation}. The presence of power-law tails in grain size distributions, arising from rare overgrowth events and kinetic fluctuations, has been further documented~\citep{pisane2015magnetic,bolyachkin2018power}, motivating the use of hybrid models such as the MLP distribution~\citep{basu2004power,basu2015mlp} to capture both typical and extreme grain size behaviors.

In magnetic materials, grain size variability has a well-established influence on coercivity, with both theoretical and computational studies demonstrating power-law scaling of coercivity with grain size under various physical regimes, and also highlighting the complex dependence of coercivity on grain boundary effects, domain wall pinning, and grain size dispersion~\citep{skomski2016size,bolyachkin2018power,herzer1990grain}. Stochastic models of coercivity in magnetic nanoparticles, such as the framework by Chakraverty \emph{et al.}~\citep{chakraverty2007coercivity}, explicitly model size-dependent switching mechanisms and the stochastic nature of magnetization reversal across single-domain and multi-domain regimes.

Stochastic modeling of magnetic hysteresis itself has advanced considerably, with approaches ranging from Preisach-type models to stochastic differential equation frameworks for domain dynamics~\citep{mayergoyz2003mathematical,bertotti1998hysteresis,yari2023static,pal2000stochastic}. Recent work emphasizes the importance of capturing both static and dynamic magnetization responses, particularly under time-varying fields~\citep{yari2023static,zidarivc2011new,gilles2002magnetic,jiles1986theory,jiles2002hysteresis}. In this context, several models have incorporated empirical grain size distributions into hysteresis simulations~\citep{bance2014grain,herzer1990grain,dutz2007hysteresis,actis2020dipolar,zhang2017magnetocaloric}, typically treating the grain size distribution as an external input.

The correlation between structural size and magnetic properties in ferrite materials has also been extensively investigated. It is well established that extrinsic properties such as coercivity are markedly influenced by grain and particle size~\citep{mahmood2018tuning}. M-type hexaferrites, in addition to their classical applications, are increasingly employed in advanced technologies including electro-acoustic devices, power tools, electronics, computing systems, and automotive components~\citep{le2022hexaferrite}. The relationship between grain size and magnetic performance is particularly important in these materials and has been widely explored~\citep{urbano2024optimization,jing2015width}.

A key concept in this context is the critical grain size, which defines the transition between single-domain and multi-domain magnetic behavior. This threshold depends on intrinsic material parameters, including domain wall energy, saturation magnetization, magnetocrystalline anisotropy, and lattice constants~\citep{luo2012physical}. As noted by Urbano-Pe{\~n}a \emph{et al.}~\citep{urbano2024optimization}, reported values for the critical grain size in hexaferrite magnets vary considerably, reflecting both synthesis conditions and measurement techniques. For barium hexaferrite, Goto \emph{et al.}~\citep{goto1980studies} estimated this size to lie between 600~nm and 1.3~$\mu$m, while Kitakami \emph{et al.}~\citep{kitakami1988study} obtained similar estimates based on magnetic domain imaging. Rezlescu \emph{et al.}~\citep{rezlescu1999fine}, using glass crystallization methods, proposed a critical size of approximately 460~nm, whereas Pullar~\citep{pullar2012hexagonal}, in a comprehensive review, reported values ranging from 300 to 900~nm depending on synthesis routes and grain growth control.

Although numerous studies have investigated the forward problem of predicting magnetic properties from structural morphology, to the best of the authors' knowledge, the inverse problem of inferring grain size statistics directly from magnetic hysteresis measurements has not yet been systematically addressed. Moreover, in most prior studies the critical grain size is either assumed based on material constants or estimated independently of magnetic measurements. In contrast, a key aspect of the present work is the treatment of the critical grain size as a latent parameter to be jointly optimized within the stochastic-dynamic hysteresis modeling framework, enabling data-driven inference of size effects directly from magnetic hysteresis loops.

\section{Stochastic Nucleation-Growth Model}
\label{sec:nucleation_growth}
We consider a general framework for modeling the statistical distribution of grain sizes resulting from thermally activated nucleation and growth in polycrystalline materials. Such processes are characteristic of a broad class of systems in which grains nucleate stochastically and evolve under both deterministic and random kinetic influences \cite{abbaschian2010physical}. The resulting microstructures typically exhibit heterogeneous grain size distributions, shaped by spatial variability in local growth conditions and stochastic interruption mechanisms \cite{porter2009phase}.

To describe this behavior, we propose a hybrid probabilistic mechanism. Grains of moderate size emerge from compounded fluctuations in thermodynamic and kinetic parameters, such as interfacial mobility, solute concentration, and defect interactions, resulting in a distribution with lognormal characteristics. Conversely, the rare but statistically significant presence of abnormally large grains often reflects delayed impingement, reduced local nucleation density, or other anomalous kinetic pathways, introducing a heavy-tailed component consistent with power-law behavior.

To unify these dual regimes, we adopt the MLP distribution \citep{basu2004power,basu2015mlp}. This model captures the multiplicative nature of early-stage nucleation and growth through a lognormal core, while accommodating rare, unbounded growth excursions via a power-law tail. The distribution arises naturally from a generative process in which each grain nucleates with a lognormally distributed initial radius and undergoes exponential growth for a random duration governed by a memoryless stopping time.

Let $\mathcal{R}_t$ denote a stochastic process representing the radius of a grain at internal kinetic time $t$. We assume that each grain nucleates at $t = t_0$ with an initial radius $\mathcal{R}_{t_0} \sim \operatorname{Lognormal}\left( \mu, \sigma^2 \right)$, reflecting heterogeneity in nucleation thermodynamics and early interface kinetics. The lognormal form arises from the multiplicative effect of random variations in local interface energies, precursor configurations, nucleation barriers, and atomic-scale factors influencing interface mobility, including the probability of finding suitable atomic positions across the interface. These combined effects introduce inherent variability in the initial effective growth rates, consistent with empirical grain size distributions observed in both experimental studies and phase-field simulations of early-stage grain formation.

\begin{figure}[!htb]
	\centering
	\includegraphics[width=1\textwidth]{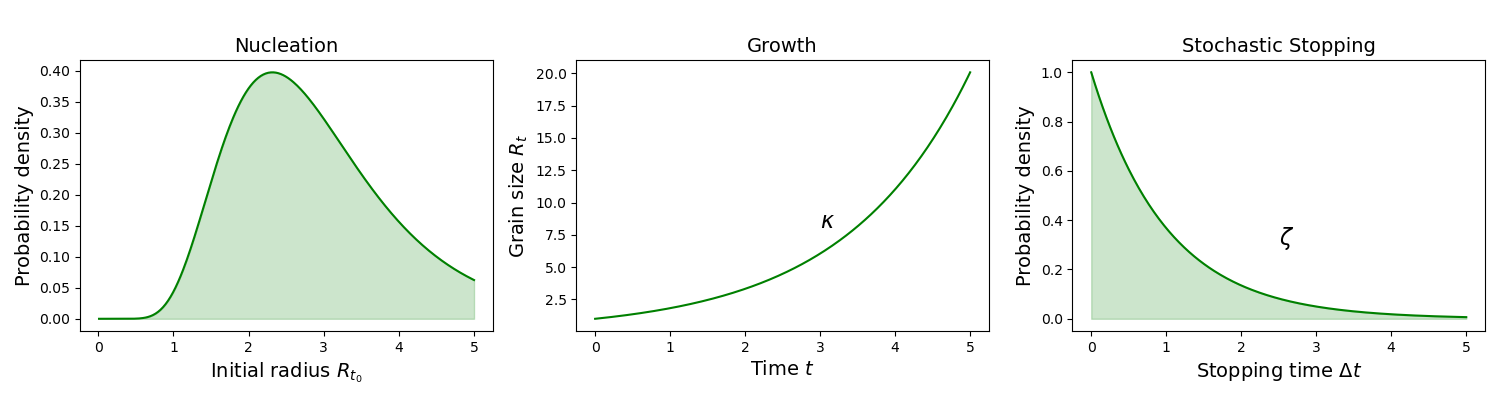}
	\caption{Schematic representation of the MLP grain formation process: (left) lognormal nucleation of initial grain radii \( \mathcal{R}_{t_0} \) parameterized by \( \mu \) and \( \sigma \); (center) deterministic exponential growth governed by rate \( \kappa \); (right) stochastic halting via exponential stopping time with rate \( \zeta \). Together, these yield the hybrid MLP distribution.}
	\label{fig:MLP_schematic}
\end{figure}

For \( t > t_0 \), grain evolution follows deterministic exponential growth governed by the ordinary differential equation
\begin{equation}
	\frac{\partial \mathcal{R}_t}{\partial t} = \kappa \mathcal{R}_t,
\end{equation}
where \( \kappa > 0 \) is the intrinsic radial \textit{growth rate}. The final grain size is not determined solely by this deterministic law, but rather through a stochastic stopping time \( \Delta t = t - t_0 \), which is modeled as exponentially distributed with density
\begin{equation}
	f_T(t) = \zeta e^{-\zeta t}, \qquad t \geq 0,
\end{equation}
where \( \zeta > 0 \) is the stochastic \textit{stopping rate}. This abstraction accommodates a variety of microscopic termination mechanisms, including intergranular impingement, solute depletion, vacancy drag, or localized energetic saturation. These effects render the lifetime of grain growth inherently random, even under nominally isothermal and homogeneous conditions. Figure \ref{fig:MLP_schematic} depicts the MLP grain formation process schematically.

The final grain radius \( R \) is then expressed as the compound random variable
\begin{equation}
	R = \mathcal{R}_{t_0} e^{\kappa \Delta t},
\end{equation}
with \( \mathcal{R}_{t_0} \sim \mathrm{Lognormal}(\mu, \sigma^2) \) and \( \Delta t \sim \mathrm{Exp}(\zeta) \), leading to a distribution governed by the MLP family, whose probability density function is given by
\begin{equation}
	g_R(r;\mu,\sigma,\omega) = \left(\dfrac{\omega}{2}\right) \left(\dfrac{1}{r}\right)^{1+\omega} \exp\left\{ \omega \mu + \dfrac{\omega^2 \sigma^2}{2} \right\} \mathrm{erfc}\left\{ \frac{1}{\sqrt{2}} \left( \omega\sigma - \frac{\log r - \mu}{\sigma} \right) \right\},
\end{equation}
with parameters $\mu \in \mathbb{R}, \sigma>0$, and $\omega=\zeta / \kappa$ denoting the \textit{tail index}. Here, $\operatorname{erfc}(\cdot)$ stands for the complementary error function. The associated cumulative distribution function is given by
\begin{equation}
	G_R(r;\mu,\sigma,\omega) = \dfrac{1}{2} \mathrm{erfc}\left\{ - \dfrac{\log r - \mu}{\sqrt{2}\sigma} \right\}  - \dfrac{1}{2r^\omega}  \exp\left\{ \omega \mu + \dfrac{\omega^2 \sigma^2}{2} \right\} \mathrm{erfc}\left\{ \dfrac{1}{\sqrt{2}} \left( \omega\sigma - \dfrac{\log r - \mu}{\sigma} \right) \right\}.
\end{equation}

The MLP distribution interpolates between lognormal and power-law behavior. As \( r \to 0 \), the lognormal core dominates, reflecting typical nucleation and bounded growth. In contrast, in the limit \( \sigma \to 0 \), the distribution approaches a pure power-law with decay exponent \( 1 + \omega \), modeling rare but prolonged growth excursions. These two asymptotic regimes capture the coexisting regularity and extremal variability observed in thermally processed polycrystalline systems.

Closed-form expressions for the first two moments of the MLP distribution exist under suitable constraints on \( \omega \). The mean and variance are given by
\begin{align}
	\mathbb{E}[R] &= \frac{\omega}{\omega - 1} \exp\left\{ \mu + \frac{\sigma^2}{2} \right\}, \qquad \omega > 1, \\
	\mathrm{Var}[R] &= \omega \exp\left\{ 2\mu + \sigma^2 \right\} \left( \frac{\exp\left\{ \sigma^2 \right\} }{\omega - 2} - \frac{\omega}{\left( \omega - 1 \right)^2} \right), \qquad \omega > 2.
\end{align}
These expressions highlight the sensitivity of the MLP distribution to the tail index \( \omega \). As \( \omega \) approaches its lower bounds, the distribution becomes increasingly heavy-tailed, and the expected value and variance diverge. This behavior underscores the importance of accurate tail modeling in systems where rare, oversized grains influence macroscopic properties such as magnetic anisotropy, intergranular stress development, or fracture initiation. In practical terms, a histogram of grain sizes illustrates these dynamics: the parameter \( \mu \) shifts the central peak, \( \sigma \) controls the dispersion, and \( \omega \) governs the steepness of the decay in the upper tail.

While the MLP model is probabilistic in structure, it retains a direct correspondence with classical recrystallization kinetics defined by the nucleation rate \( N \) and the growth rate \( G \). In traditional kinetic theory, high \( N \) and low \( G \) produce many small grains due to rapid nucleation and slow expansion, resulting in narrow grain size distributions. In contrast, low \( N \) and high \( G \) yield a few dominant grains and a broader distribution \cite{fredriksson2012solidification}. These macroscopic behaviors are encoded in the MLP parameters: \( \mu \) and \( \sigma \) describe the typical scale and heterogeneity of initial conditions, while \( \omega = \zeta / \kappa \) governs the likelihood of extreme grain sizes and can be viewed as a statistical surrogate for the effective \( N/G \) ratio. Low \( \omega \) corresponds to tail-heavy regimes dominated by growth kinetics, while high \( \omega \) signals tightly constrained, nucleation-controlled evolution.

The MLP model is particularly suitable for systems undergoing unconstrained or semi-constrained growth, where grains evolve independently and termination is driven by local stochastic effects. It is less applicable to systems characterized by long-range spatial correlations, orientation-dependent impingement, or externally imposed anisotropic fields, where the growth dynamics cannot be reduced to localized interactions and memoryless halting.

Hence, the MLP framework provides a statistically expressive, physically grounded model for grain size evolution in thermally activated systems. It generalizes classical nucleation-growth kinetics, accommodates heterogeneous and rare-growth dynamics, and supports empirical calibration and predictive modeling across a wide range of polycrystalline materials.

\section{Stochastic Hysteresis Modeling}
\label{sec:magnetic_modeling}
The magnetic response of polycrystalline materials emerges from complex interactions between grain-scale structures and external fields \cite{buschow2003handbook}. To model this behavior, we combine a stochastic description of coercivity, rooted in grain size statistics, with a dynamical system for time-dependent magnetization. The approach accounts for both static heterogeneity and viscous or inertial effects that lead to hysteresis. In what follows, we first derive a probability distribution for coercivity from a latent grain radius model. We then couple this distribution to a first-order magnetization dynamic, yielding a unified framework capable of reproducing rate-dependent hysteresis loops and inferring microstructural parameters from empirical data.

\subsection{Coercivity Distribution}

We now extend the grain size distribution framework to model magnetic properties stochastically, by linking the physical behavior of coercivity to the underlying statistical distribution of grain radii. Magnetic coercivity, denoted by \( H \), is known to depend sensitively on grain size, particularly in systems where magnetic reversal processes are influenced by domain wall pinning or coherent rotation mechanisms \cite{singh2023ferrite}. One classical model capturing this dependence is Brown's relation, which approximates coercivity in terms of grain radius \( R \) by the following inverse-linear transformation
\begin{equation}
	\label{Brown}
	H = H_0 \left(1 - \frac{R_c}{R} \right), \qquad R > R_c,
\end{equation}
where \( H_0 \) is the theoretical coercivity for ideal single-domain grains and \( R_c \) is the critical radius below which coherent rotation dominates and above which domain wall motion begins to reduce coercivity. This relation is monotonic and differentiable on the domain \( R > R_c \), allowing us to construct a stochastic model for coercivity by propagating the MLP-distributed radius \( R \) through the transformation \eqref{Brown}.

To derive the probability density function of coercivity, we apply the standard Jacobian method for transformed random variables. Let \( R \sim \mathrm{MLP}(\mu, \sigma, \omega) \) with density function \( g_R(r;\mu,\sigma,\omega) \). The transformation \eqref{Brown} is invertible on the range \( h \in (0, H_0) \), and its inverse is given by
\begin{equation}
	R = \frac{R_c H_0}{H_0 - H}.
\end{equation}
Differentiating with respect to \( h \), we obtain the Jacobian
\begin{equation}
	\frac{dr}{dh} = \frac{R_c H_0}{(H_0 - h)^2}.
\end{equation}
The coercivity probability density function is then given by
\begin{equation}
	f_H(h) = g_R\left( \frac{R_c H_0}{H_0 - h};\, \mu, \sigma, \omega \right) \cdot \frac{R_c H_0}{(H_0 - h)^2},
	\qquad h \in (0, H_0).
\end{equation}
Substituting the explicit form of the MLP density, the coercivity density function becomes
\begin{align}
	f_H(h) = & \left(\dfrac{\omega}{2}\right) \left( \frac{1}{H_0 - h} \right)^{1 - \omega}
	\left( \frac{1}{R_c} \right)^{1+\omega}
	\exp\left\{ \omega \mu + \dfrac{\omega^2 \sigma^2}{2} \right\} \nonumber \\
	& \times \mathrm{erfc}\left\{ \frac{1}{\sqrt{2}} \left( \omega\sigma - \frac{\log \left( \dfrac{R_c H_0}{H_0 - h} \right) - \mu}{\sigma} \right) \right\}, \qquad h \in (0, H_0).
\end{align}

We now derive a closed-form expression for the mean coercivity. Using the transformation \eqref{Brown}, we write
\begin{equation}
	\mathbb{E}[H] = H_0 \left(1 - R_c\, \mathbb{E}\left[ \frac{1}{R} \right] \right).
\end{equation}
From the moment-generating properties of the MLP distribution \cite{basu2004power, basu2015mlp}, the inverse moment is given by
\begin{equation}
	\mathbb{E}\left[ \frac{1}{R} \right] = \frac{\omega}{\omega + 1} \exp\left( -\mu + \frac{\sigma^2}{2} \right).
\end{equation}
Therefore, the expected coercivity is
\begin{equation}
	\mathbb{E}[H] = H_0 \left(1 - R_c\, \frac{\omega}{\omega + 1} \exp\left( -\mu + \frac{\sigma^2}{2} \right) \right),
\end{equation}
which expresses the mean coercive field directly in terms of the MLP parameters. This result quantifies how the average magnetic response depends jointly on nucleation scale (\( \mu \)), heterogeneity (\( \sigma \)), and the balance between growth and stochastic arrest (\( \omega \)).

In practice, however, only grains larger than the critical radius \( R_c \) are physically meaningful under the Brown model. To respect this domain constraint, we define the conditional coercivity distribution
\begin{equation}
	f_{H \mid R > R_c}(h) = \frac{f_H(h)}{1 - G_R(R_c; \mu, \sigma, \omega)},
\end{equation}
where \( G_R(R_c; \mu, \sigma, \omega) \) is the cumulative distribution function of the MLP evaluated at \( R = R_c \). This conditional distribution properly accounts for the truncated support of the underlying radius process and ensures that coercivity statistics are computed over the physically relevant regime. It refines the theoretical model for applications where precise alignment with micromagnetic principles is required.

In this study, rather than estimating \( \mu, \sigma, \omega \) and \( R_c \) independently from microstructural data, we calibrate them directly from the full magnetization loop. The coercivity distribution \( f_H(h) \) is not postulated independently, but inferred implicitly through joint optimization over the hysteresis response. The critical radius \( R_c \), typically regarded as a fixed micromagnetic threshold, is treated as a fit parameter that governs the inflection scale of the coercivity function. This enables the data to determine the effective boundary between coherent rotation and domain-wall mediated switching regimes within the heterogeneous grain population.

\subsection{Dynamic Magnetization}

We further extend the coercivity-based framework to a dynamic time-resolved model of magnetization by introducing explicit time dependence in the applied field. Let \( H(t) \) denote a prescribed, time-varying external magnetic field. In practical settings, such as AC magnetic measurements or cyclic magnetization testing, the field is often periodic; i.e.,
\begin{equation}
	H(t) = H_{\mathrm{max}} \sin(2\pi f t),
\end{equation}
where \( f \) is the cycling frequency and \( H_{\mathrm{max}} \) is the field amplitude.

For a material composed of an ensemble of grains with distributed coercivities governed by the density \( f_H(h) \), we define the instantaneous equilibrium magnetization at time \( t \) as follows:
\begin{equation}
	M_{\mathrm{eq}}(t) = M_s \int_0^{H_0} \tanh\left( \frac{H(t)}{h} \right) f_{H \mid R > R_c}(h) \, dh.
\end{equation}
This quantity captures the ideal magnetization state under quasi-static conditions, assuming instantaneous alignment of grains according to their respective reversal thresholds.

However, real materials exhibit delayed response to field changes due to domain wall inertia, eddy currents, or thermal activation barriers. To incorporate this finite response time, we introduce a first-order relaxation model as follows:
\begin{equation}
	\tau \frac{dM(t)}{dt} + M(t) = M_{\mathrm{eq}}(t),
\end{equation}
where \( \tau \) is the effective \textit{magnetization relaxation time}. This equation accounts for the rate-limited adjustment of the system toward the instantaneous equilibrium state. It is formally equivalent to a low-pass temporal filter applied to the equilibrium signal and smooths rapid changes in \( M(t) \) under fast cycling or high field gradients.

\begin{figure}
	\centering
	\includegraphics[scale=0.4]{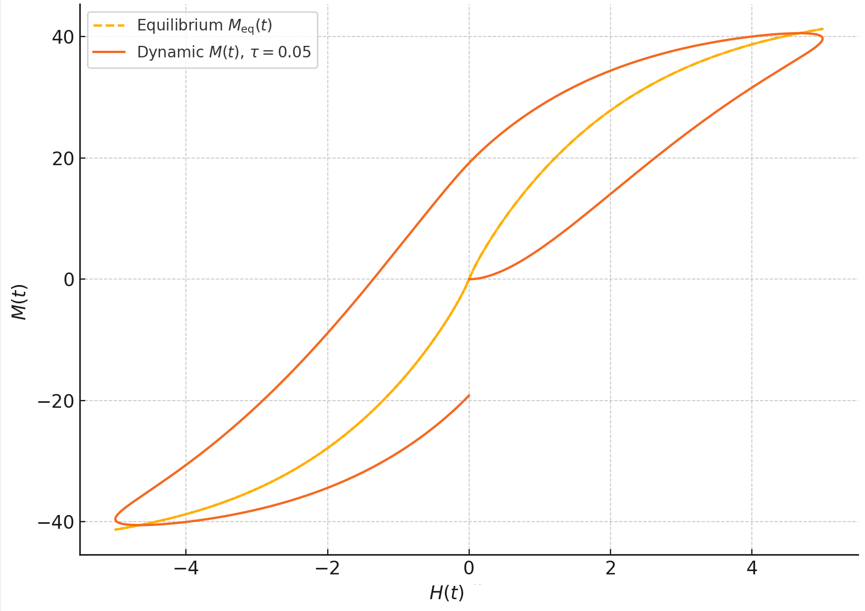}
	\caption{Schematic of dynamic hysteresis with viscous lag.}
	\label{hysterisis}
\end{figure}

Solving this equation numerically over one or more periods of \( H(t) \) yields the time-resolved magnetization response \( M(t) \), which when plotted against \( H(t) \) produces a dynamic hysteresis loop; see e.g., Figure \ref{hysterisis}. This loop naturally exhibits broader tails and greater area at higher frequencies or larger \( \tau \), consistent with experimental observations of rate-dependent hysteresis loss.

This framework therefore provides a unified stochastic, dynamic model: the coercivity distribution encodes grain-level disorder, while the dynamic ordinary differential equation captures macroscopic lag, together explaining both static and dynamic features of observed magnetic hysteresis.

In our formulation, the relaxation time \( \tau \) is treated as an additional parameter to be identified from data, allowing the magnetization model to adapt to material-specific viscous, inertial, or eddy-current effects. The full model, comprising \( (\mu, \sigma, \omega, R_c, \tau) \), is calibrated directly by minimizing the discrepancy between simulated and measured hysteresis loops. This inversion yields not only predictive magnetization dynamics, but also interpretable microstructural and kinetic parameters rooted in grain-scale physics.

\section{Experiments and Discussion}
\label{sec:experiments}
To validate the stochastic-dynamic model of magnetic hysteresis developed in this work, we perform parameter estimation using empirical magnetization data across three distinct processing stages of M-type strontium hexaferrite synthesized hydrothermally. Unlike traditional approaches that rely on microstructural imaging to estimate grain size distributions, our method infers the parameters of the MLP distribution directly from hysteresis behavior, using the bulk magnetic response as a statistical proxy for the underlying grain ensemble.

The parameter estimation problem is formulated as an inverse model calibration: given empirical magnetization data \( \{H_i, M(H_i)\}_{i=1}^n \), we solve for the parameter vector \( \boldsymbol{\theta} = (\mu, \sigma, \omega, \tau, R_c) \) that minimizes the discrepancy between measured and predicted magnetization. The loss function is expressed as follows:
\begin{equation}
	\mathcal{L}(\boldsymbol{\theta}) = \sum_{i=1}^{n} \left( M(H_i; \boldsymbol{\theta}) - M(H_i) \right)^2,
\end{equation}
where \( M(H_i; \boldsymbol{\theta}) \) is computed via a conditional MLP coercivity distribution coupled with a first-order relaxation model for dynamic magnetization.

This methodology circumvents common pitfalls in grain segmentation from microscopy imaging techniques such as scanning electron microscopy (SEM) and transmission electron microscopy (TEM). These techniques often suffer from challenges including overlapping grains, limited depth of field, and insufficient sampling area. Particularly at the nano- and micro-scale, individual SEM or TEM images rarely provide statistically representative insight into the full distribution of grain sizes. By contrast, hysteresis loops encode ensemble-level signatures of coercivity distribution and magnetic lag, making them powerful candidates for indirect microstructural inference.

The experimental analysis was conducted on three material states. The raw sample was hydrothermally synthesized at \(220^\circ\mathrm{C}\) for 1 hour in the presence of NaOH, then calcined at \(850^\circ\mathrm{C}\) for 2 hours in air. The nitrogen-treated sample underwent thermal decomposition under a static nitrogen atmosphere at temperatures exceeding \(500^\circ\mathrm{C}\), followed by the same calcination step; see \citep{ataie1996heat} for further details. The final calcined sample represents the post-treatment reconstitution of the ferrite phase by exhibiting the so-called structural memory effect \citep{Molaei2025}.

The estimated parameters for each sample are summarized in Table~\ref{tab:params_experiment}. We emphasize that unlike prior studies, the critical radius \( R_c \) is not fixed but treated as a free parameter constrained within a physically admissible range, thus offering additional flexibility and precision in characterizing coercivity transitions.

\begin{table}[ht]
	\centering
	\caption{Optimal parameters inferred from magnetization data.}
	\label{tab:params_experiment}
	\begin{tabular}{|c|c|c|c|c|c|c|c|}
		\hline
		\textbf{Sample} & \(\mu\) & \(\sigma\) & \(\omega\) & \(\tau\) (s) & \(R_c\) (\(\mu\)m) & \(\mathbb{E}[R]\) (\(\mu\)m) & \(\mathbb{E}[H]\) (kOe) \\
		\hline
		Raw       & 6.147 & 0.339 & 3.034 & 0.534 & 0.608 & 0.737 & 2.763 \\
		Treatment & 5.039 & 0.770 & 3.321 & 0.344 & 0.249 & 0.297 & 0.817 \\
		Calcined  & 5.124 & 1.381 & 3.588 & 0.838 & 0.581 & 0.604 & 2.190 \\
		\hline
	\end{tabular}
\end{table}
In the raw hydrothermal sample, the high \(\mu = 6.147\) and low \(\sigma = 0.339\) indicate a well-developed and relatively uniform grain ensemble. The moderate tail index \(\omega = 3.034\) reflects a microstructure where growth is somewhat bounded but still allows occasional excursions into larger grain sizes. The critical radius \( R_c = 0.608 \,\mu\mathrm{m} \) is consistent with coherent reversal mechanisms dominating over domain-wall motion in much of the particle population. A moderately high relaxation time \( \tau = 0.534\,\mathrm{s} \) suggests moderate inertia in magnetization dynamics, potentially due to single-domain behavior.

Following nitrogen treatment, the system exhibits dramatic grain fragmentation and microstructural disorder \citep{ataie1996heat}. The drop in \(\mu = 5.039\) and increase in \(\sigma = 0.770\) reflect grain breakup and local variability. The increase in \(\omega = 3.321\) implies growth suppression, consistent with halted grain expansion from thermochemical decomposition. The low \(R_c = 0.249\,\mu\mathrm{m}\) further supports a refinement of the critical magnetic domain scale. These trends align with prior studies that attribute nitrogen treatment to the breakdown of the hexaferrite lattice and formation of finer magnetic grains \citep{ataie1996heat}. The fast relaxation time \( \tau = 0.344\,\mathrm{s} \) suggests that magnetization responds quickly to field changes, likely due to reduced anisotropy and pinning.

\begin{figure}[ht]
	\centering
	\includegraphics[width=0.55\textwidth]{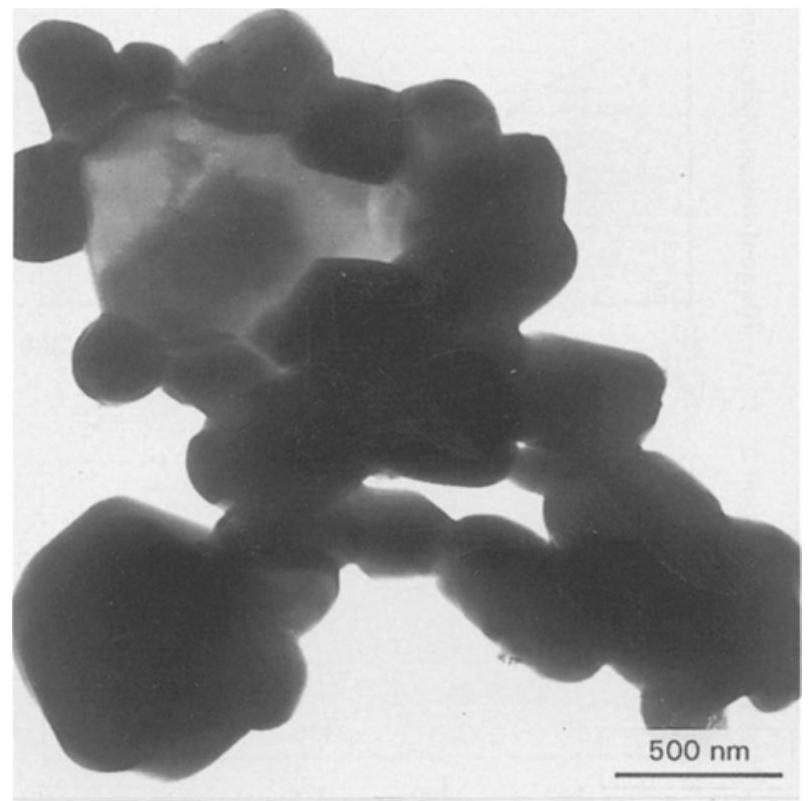}
	\caption{Transmission electron micrograph of SrFe\textsubscript{12}O\textsubscript{19} powder after nitrogen treatment and calcination, showing polycrystalline nature of particles with internal subgrains \citep{ataie1996heat}.}
	\label{fig:tem_subgrains}
\end{figure}

\begin{figure}[ht]
	\centering
	\includegraphics[width=0.5\textwidth]{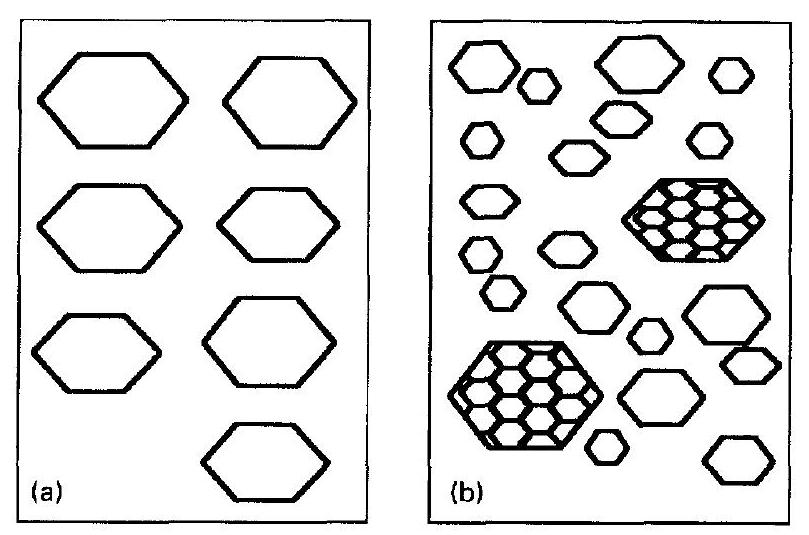}
	\caption{Schematic diagram showing transformation from (a) single-crystal particles to (b) multicrystalline and subgrain structures after treatment and calcination \citep{ataie1996heat}.}
	\label{fig:schematic_subgrain}
\end{figure}

Post-calcination, the MLP parameters exhibit partial recovery. The value \(\mu = 5.124\) increases while \(\sigma = 1.381\) expands, capturing stochastic nucleation and heterogeneous growth during phase reformation. The high \(\omega = 3.588\) indicates tightly bounded grain expansion, consistent with strong spatial constraints during recrystallization. These constraints are physically visible in the transmission electron micrograph shown in Figure~\ref{fig:tem_subgrains}, where large particles exhibit polycrystalline interiors composed of smaller subgrains. The schematic diagram in Figure~\ref{fig:schematic_subgrain} further illustrates how initial single-crystal grains are subdivided into multicrystalline and fine-grained components, preserving the outer morphology but modifying internal structure.

This behavior exemplifies the concept of structural memory: although the internal grain arrangement is reset, the outer geometry persists, imposing growth boundaries on the recrystallizing grains. In our model, this memory is encoded in the tail index \(\omega\), which suppresses rare excursions into large radii. The post-calcination value \(\omega = 3.588\) thus reflects a tight statistical envelope imposed by prior particle morphology.

The critical radius \(R_c\), typically taken as a fixed micromagnetic scale, is in this work treated as an optimization parameter. Across samples, \(R_c\) reflects how microstructural transitions influence the threshold between coherent rotation and domain-wall reversal. The decrease from \(R_c = 0.608\,\mu\mathrm{m}\) in the raw state to \(0.249\,\mu\mathrm{m}\) after treatment, and the partial recovery to \(0.581\,\mu\mathrm{m}\) post-calcination, supports this interpretation. The inferred values reflect how magnetization response becomes spatially constrained and later partially relaxed.

\begin{figure}[!htb]
	\centering
	\begin{subfigure}[b]{0.5\textwidth}
		\centering
		\includegraphics[width=\textwidth]{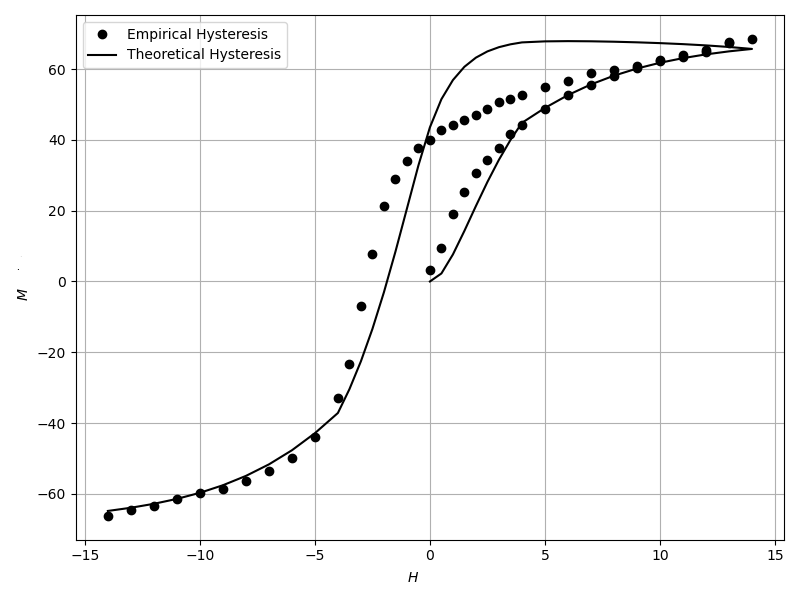}
		\caption{Raw powder.}
	\end{subfigure}
	
	\vspace{0.1cm}
	
	\begin{subfigure}[b]{0.5\textwidth}
		\centering
		\includegraphics[width=\textwidth]{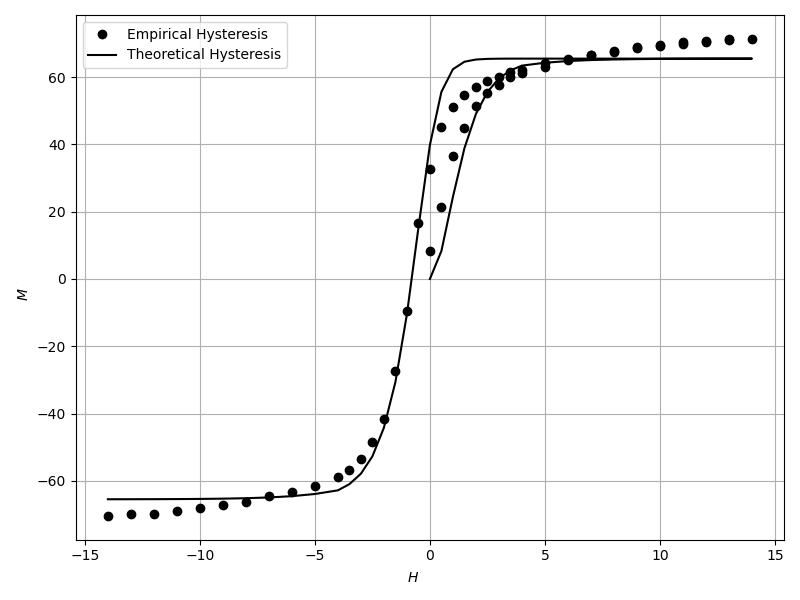}
		\caption{Nitrogen-treated powder.}
	\end{subfigure}
	
	\vspace{0.1cm}
	
	\begin{subfigure}[b]{0.5\textwidth}
		\centering
		\includegraphics[width=\textwidth]{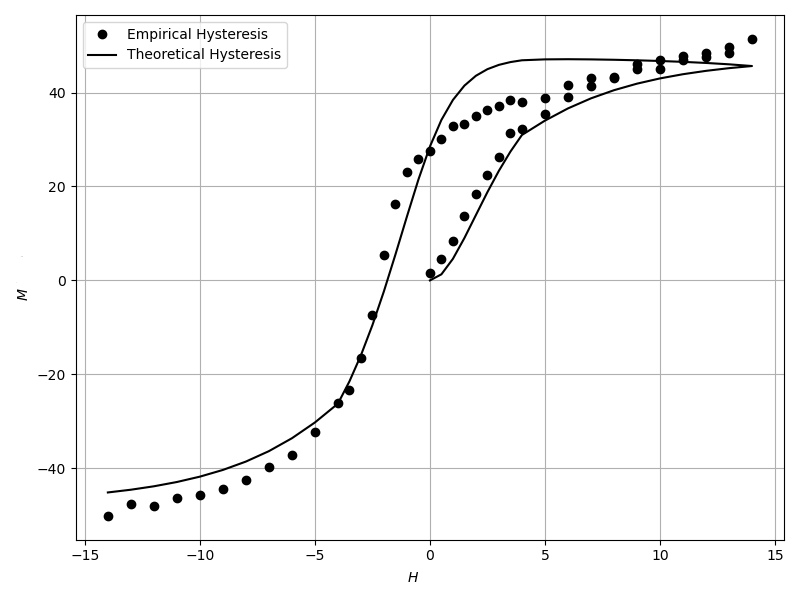}
		\caption{Calcined powder.}
	\end{subfigure}
	
	\caption{Empirical (dots) and predicted (line) hysteresis loops. The external magnetic field is reported in kOe, and magnetization is expressed in emu/g.}
	\label{fig:hysteresis}
\end{figure}

Figure \ref{fig:hysteresis} compares the empirical hysteresis loops with theoretical predictions. The model accurately recovers coercivity, slope, and saturation in each case. Minor discrepancies near the reversal regions may stem from physical effects not yet modeled, including local anisotropy and domain interaction.

Overall, these results confirm that the proposed coercivity and magnetization model offers a physically interpretable, quantitatively accurate, and experimentally verifiable pathway to infer grain-scale properties from macroscopic hysteresis. It establishes a foundation for statistical microstructural analysis in ferrites and other functional ceramics using bulk magnetic data.

\newpage
\section{Conclusion}
\label{sec:conclusion}

We have developed and validated a stochastic-dynamic framework that links grain-scale microstructural variability to magnetic hysteresis behavior in polycrystalline hexaferrite materials. By formulating a nucleation-growth model governed by probabilistic kinetics, we derived the Modified Lognormal Power-law distribution as a physically interpretable statistical descriptor of grain size evolution. This distribution was propagated through Brown's relation to produce a stochastic coercivity distribution, and subsequently embedded within a dynamic magnetization model incorporating first-order relaxation dynamics.

A central feature of this work is the joint estimation of key microstructural parameters, including the critical grain radius, directly from hysteresis measurements. This inversion approach circumvents limitations of microscopy-based grain segmentation and enables reliable parameter identification even in the presence of nanoscale heterogeneity and statistical undersampling. In doing so, the model not only predicts hysteresis curves with high fidelity but also reveals meaningful structural descriptors such as the nucleation scale, dispersion, growth-stopping kinetics, and viscous lag.

Application to hydrothermally synthesized strontium hexaferrite across multiple processing stages, raw, nitrogen-treated, and post-calcination, demonstrated the capacity of the model to track kinetic transformations and structural evolution. In particular, the estimated tail index provided a quantitative lens on the competition between nucleation and growth, reflecting the degree of grain size dispersion and the suppression or amplification of anomalous grains. The recovery of grain structure following calcination, despite prior decomposition, was captured statistically through the inferred narrowing of the MLP distribution and increase in effective coercivity, offering a quantitative interpretation of the structural memory effect.

This methodology establishes hysteresis as a viable probe of latent microstructural states. It opens new possibilities for material diagnostics, optimization of processing protocols, and inverse design of magnetic materials based on their dynamic response. Future work will extend the framework to incorporate spatial heterogeneity, field-history dependence, and multi-modal grain populations, as well as generalize the inversion to include frequency-dependent loss measurements and multi-dimensional magnetization processes.



\end{document}